\documentclass[structabstract]{aa} 
\usepackage{graphicx}
\usepackage[varg]{txfonts}
\usepackage{natbib}
\usepackage{txfonts}
\usepackage{fixltx2e}
\usepackage{color}
\usepackage{enumerate}
\usepackage{amstext}
\usepackage{xspace}
\usepackage{pdfpages}
\bibpunct{(}{)}{;}{a}{}{,}
\newcommand{\scname}[1]{\textsc{#1}}
\newcommand{\snf}{\scname{SNfactory}\xspace}
\usepackage{epstopdf}
\begin{document}
\title{Measuring cosmic bulk flows with Type Ia Supernovae from the
  Nearby Supernova Factory}

\authorrunning{U.~Feindt, M.~Kerschhaggl, M.~Kowalski \& the \snf}

\newcounter{savecntr}
\newcounter{restorecntr}
\author
{
    U.~Feindt\inst{1},
    M.~Kerschhaggl\inst{1},
    M.~Kowalski\inst{1},
    G.~Aldering\inst{2},
    P.~Antilogus\inst{3},
    C.~Aragon\inst{2},
    S.~Bailey\inst{2},
    C.~Baltay\inst{4},
    S.~Bongard\inst{3},
    C.~Buton\inst{1},
    A.~Canto\inst{3},
    F.~Cellier-Holzem\inst{3},
    M.~Childress\inst{5},
    N.~Chotard\inst{6},
    Y.~Copin\inst{6},
    H.~K. Fakhouri\inst{2,7},
    E.~Gangler\inst{6},
    J.~Guy\inst{3}, 
    A.~Kim\inst{2},
    P.~Nugent\inst{8,9},
    J.~Nordin\inst{2,10},
    K.~Paech\inst{1},
    R.~Pain\inst{3},
    E.~Pecontal\inst{11},
    R.~Pereira\inst{6},
    S.~Perlmutter\inst{2,7},
    D.~Rabinowitz\inst{4},
    M.~Rigault\inst{6},  
    K.~Runge\inst{2},
    C.~Saunders\inst{2},
    R.~Scalzo\inst{5},
    G.~Smadja\inst{6},
    C.~Tao\inst{12,13},
    R.~C.~Thomas\inst{8},
    B.~A.~Weaver\inst{14},
    C.~Wu\inst{3,15}
}

\institute{
    Physikalisches Institut, Universit\"at Bonn,
    Nu\ss allee 12, 53115 Bonn, Germany\\
    e-mail: \texttt{[feindt,mkersch]@physik.uni-bonn.de}
\and
    Physics Division, Lawrence Berkeley National Laboratory, 
    1 Cyclotron Road, Berkeley, CA, 94720
\and
    Laboratoire de Physique Nucl\'eaire et des Hautes \'Energies,
    Universit\'e Pierre et Marie Curie Paris 6, Universit\'e Paris Diderot Paris 7, CNRS-IN2P3, 
    4 place Jussieu, 75252 Paris Cedex 05, France
\and
    Department of Physics, Yale University, 
    New Haven, CT, 06250-8121
\and
    Research School of Astronomy and Astrophysics, Australian National
    University, Canberra, ACT 2611, Australia.
\and
    Universit\'e de Lyon, F-69622, Lyon, France ; Universit\'e de Lyon 1, Villeurbanne ; 
    CNRS/IN2P3, Institut de Physique Nucl\'eaire de Lyon.
\and
    Department of Physics, University of California Berkeley,
    366 LeConte Hall MC 7300, Berkeley, CA, 94720-7300
\and
    Computational Cosmology Center, Computational Research Division, Lawrence Berkeley National Laboratory, 
    1 Cyclotron Road MS 50B-4206, Berkeley, CA, 94720
\and
    Department of Astronomy, B-20 Hearst Field Annex \# 3411,
    University of California, Berkeley, CA 94720-3411, USA
\and
    Space Sciences Laboratory, University of California Berkeley, 7 Gauss
    Way, Berkeley, CA 94720, USA
\and
    Centre de Recherche Astronomique de Lyon, Universit\'e Lyon 1,
    9 Avenue Charles Andr\'e, 69561 Saint Genis Laval Cedex, France
\and
    Centre de Physique des Particules de Marseille, 163, avenue de Luminy - Case 902 - 13288 Marseille Cedex 09, France
\and
    Tsinghua Center for Astrophysics, Tsinghua University, Beijing 100084, China 
\and
    Center for Cosmology and Particle Physics,
    New York University,
    4 Washington Place, New York, NY 10003, USA
\and
    National Astronomical Observatories, Chinese Academy of Sciences, Beijing 100012, China
}

\date{Received 12 May 2013 / Accepted 10 Oct 2013}

 
   \abstract
   {Our Local Group of galaxies appears to be moving relative to the
     Cosmic Microwave Background with the source of the peculiar
     motion still uncertain. While in the past this has been studied
     mostly using galaxies as distance indicators, the weight of
     type~Ia supernovae (SNe~Ia) has increased recently with the continuously improving
     statistics of available low-redshift supernovae.}
   {We measured the bulk flow in the nearby universe ($0.015<z<0.1$)
     using 117 SNe~Ia observed by the Nearby Supernova Factory, as
     well as the Union2 compilation of SN~Ia data already in the
     literature.}
   {The bulk flow velocity was determined from SN data binned in
     redshift shells by including a coherent motion (dipole) in a
     cosmological fit. Additionally, a method of spatially smoothing
     the Hubble residuals was used to verify the results of the dipole
     fit. To constrain the location and mass of a potential mass
     concentration (e.g.\ the Shapley Supercluster) responsible for
     the peculiar motion, we fit a Hubble law modified by adding an
     additional mass concentration.}
   {The analysis shows a bulk flow consistent with the direction of
     the CMB dipole up to $z\sim 0.06$, thereby doubling the volume
     over which conventional distance measures have sensitivity to a
     bulk flow. We see no significant turnover behind the center of
     the Shapley Supercluster. A simple attractor model in the
     proximity of the Shapley Supercluster is only marginally
     consistent with our data, suggesting the need for another, more
     distant, source. In the redshift shell $0.06<z<0.1$, we constrain
     the bulk flow velocity to $\leq240$~km~s$^{-1}$ (68\% confidence
     level) for the direction of the CMB dipole, in contradiction to
     recent claims of the existence of a large amplitude dark flow.}
   {}

 \titlerunning{Measuring cosmic bulk flows}
   \maketitle
%

\section{Introduction}
The Copernican principle, implying an isotropic Universe on large
scales, is one of the major conceptual building blocks of modern
cosmology. In this picture, an important task is to explain the
apparent motion of our Local Group of galaxies (LG) relative to the
Cosmic Microwave Background \citep{kogut} with
$v_{\mathrm{LG}}=627\pm22$~km~s$^{-1}$ toward the direction
$l=276^\circ$ and $b=30^\circ$. While the gravitational attraction
towards a nearby overdensity is widely accepted as the source of the
LG motion, the exact contribution of known overdensities is still
under debate as the amplitude of $v_{\mathrm{LG}}$ has not been fully
recovered by peculiar velocity measurements in the local universe. At
large scales -- where the peculiar velocity data become sparser and
noisier, thus precluding the reconstruction of a full peculiar
velocity field -- the mass distribution can be investigated by
measuring bulk flows, i.e.\ coherent motion in large volumes.

Previous studies of the local bulk flow show possible tension between
two sets of results: some analyses reported possible anomalously large
bulk flows at scales of $\sim100\,h^{-1}$~Mpc
\citep{hudson04,watkins,lavaux,colin,macaulay} while others find the
bulk flow to be consistent with the expectation from $\Lambda$CDM
(earliest \citealt{courteau}, more recently
\citealt{nusser,nusser11b,branchini12,turnbull,ma13}). \cite{lavaux},
for example, found in the 2MRS galaxy catalog that the LG velocity
$v_{\mathrm{LG}}$ and the direction towards the CMB dipole could not
be recovered for matter restframes at distances less than
$120\,h^{-1}$ Mpc. On the other hand, \citeauthor{nusser11b} showed
for the same catalog and for the SF++ galaxy catalog that the bulk
flow amplitude is consistent with $\Lambda$CDM
expectations~\citep{nusser11b,nusser}.

Over larger distances, \cite{kashlinsky08} reported a strong and
coherent bulk flow out to $d\gtrsim 300\,h^{-1}\textrm{Mpc}$ based on
measurements of the kinematic Sunyaev-Zeldovich effect of X-ray
clusters. According to their latest results (\citeauthor{kashlinsky08}
\citeyear{kashlinsky08,kashlinsky09,kashlinsky10,kashlinsky11,kashlinsky12})
the bulk flow is $\sim$1000~km~s$^{-1}$ in the direction of the CMB
dipole up to a distance of at least $\sim$800~Mpc. The large scale
structure formation predicted by the $\Lambda$CDM model does not
explain such large values for bulk flows at these distances. A
possible explanation for this ``Dark Flow'' is a tilt imprinted on our
horizon by pre-inflationary inhomogeneities
\citep{grishchuk78,turner91}. However, this claim has been questioned
in various studies \citep{keisler09,osborne11,mody12,lavaux13} and
recently possibly rejected by the
\cite{planck_xiii}.\footnote{However, \cite{atrio13} argued that the
  uncertainties were overestimated in this study and hence the bulk
  flow measured by Planck is more significant than reported and
  consistent with the previous results of \cite{kashlinsky10}.}

At smaller distances, the mass distribution can be studied more
precisely by reconstructing the peculiar velocity field using large
galaxy catalogs \citep{erdogdu2006,lavaux}. The contribution of
overdensities at those scales to the LG motion has long been studied.
\cite{lynden} measured a flow that suggested a \emph{Great Attractor}
(GA) at about $43\,h^{-1}$ Mpc. However, \cite{kocevski} found that
the GA only accounts for $44\,\%$ of the dipole anisotropy in a large
X-ray cluster sample, with the rest evidently caused by more distant
sources such as the Shapley Supercluster (SSC) at a distance of
$\textrm{105-165}\,h^{-1}~\textrm{Mpc}$ ($0.035<z<0.055$) in the
direction $l=306.44^{\circ}, b=29.71^{\circ}$
\citep{shapley30,scaramella89,raychaudhury91}. In addition to
attraction due to nearby overdensities, outflows from local voids
contribute to the LG motion. \citet{tully08} decomposed the peculiar
motion of the Local Sheet in which the LG is embedded into three
almost orthogonal components. Two of these are attributed to
attraction by the Virgo and Centaurus clusters while the other is a
velocity of 259~km~s$^{-1}$ toward $l=209.7^{\circ},b=-2.6^{\circ}$,
away from the Local Void. Due to the increasing sparseness of data on
larger scales it is still not clear whether convergence of
$v_{\mathrm{LG}}$ can be achieved including the SSC or if a
gravitational source extending even further out is responsible for the
CMB dipole \citep{lavaux}. Hence the study of bulk flows at large
distances will yield valuable information on the source of the LG
motion.

When determining the velocity of the LG, other effects that contribute
to the CMB dipole need to be considered as well. \cite{wiltshire}
found in an analysis of the Hubble flow variance of 4534 galaxy
distances that the mean of the Hubble parameter of consecutive
redshift shells is more compatible with its global value when using
the Local Group rather than the CMB as rest frame. In addition to a
local boost it is suggested that the CMB dipole could be attributed to
differences in the apparent distance to the surface of last scattering
mediated by foreground structures at the $60\,h^{-1}$ Mpc scale.

Peculiar velocity fields in the nearby Universe have long been
investigated using galaxies as distance indicators, where the distance
is estimated through the Tully-Fisher relation \citep{tully77}, the
Fundamental Plane \citep{djorgovski87,dressler87} or Surface
Brightness fluctuations \citep{tonry88}, see e.g. \cite{watkins} and
\cite{lavaux} and references therein. However, galaxies as distance
indicators generally range out around $100\,h^{-1}$~Mpc as the
accuracy per galaxy becomes low and the observational cost large.
Supernovae of Type~Ia (SNe~Ia), on the other hand, are bright
standardizable candles, i.e. showing a brightness dispersion of
$\sim$10\% after empirical corrections \citep{phillips}, and thus are
alternative tracers of bulk flow motions exceeding by far the redshift
range of galaxy distance indicators. While brightness and
standardizability favor SNe~Ia, the lack of large samples has limited
their use to only a few studies so far. Early studies of nearby SN
data showed that the motion of the LG is consistent with the measured
dipole seen in the Cosmic Microwave Background CMB \citep{riess}, and
this has been confirmed a number of times since then
\citep{2007ApJ...661..650H,Gordon:2007zw,Gordon:2007sk,2011ApJ...732...65W}.
One of the most recent studies of a dataset of 245 nearby SNe~Ia
resulted in a bulk flow towards $l=319^{\circ}\pm18^{\circ}$,
$b=7^{\circ}\pm14^{\circ}$ at a rate of
$249\pm76\,\mbox{km}\,\mbox{s}^{-1}$ \citep{turnbull}. Moreover,
\citet{schwarz} discovered a statistically significant hemispheric
anisotropy at $>95\,\%$ confidence level in several SN~Ia datasets,
for SNe at $z<0.2$. They found an asymmetry between the north and
south equatorial hemispheres. \citet{2012arXiv1212.3691K} recently
verified this with a larger dataset that the asymmetry does not
contradict $\Lambda$CDM expectations.

Since the coordinates for the GA, SSC and CMB dipole are compatible
across the existing studies, a clarification of the situation
described above can be approached through the study of redshift
shells. Recently \cite{colin} investigated anisotropies in discrete
redshift shells using the Union2 compilation of Type~Ia SNe
\citep{amanullah}. This \emph{z-tomography} apparently revealed a SSC
signature imprinted on the nearby Hubble flow. However, this existing
SN~Ia sample lacks sufficient variance-weighted depth. Hence new
data are needed to clarify the nature of nearby bulk flows.

In this paper we present an anisotropy study using new SN distance
measurements from the Nearby Supernova Factory (\snf,
\citealt{aldering}), a project that focuses on a large set of
spectrophotometrically observed SNe~Ia in the redshift range
$0.03<z<0.08$ ($ \textrm{90-240}~h^{-1}$ Mpc). The \snf dataset used
here consists of 117 SNe~Ia and more than doubles the number of
distance measurements in this redshift range and provides --- for the
first time with conventional distance measurements --- sensitivity to
a bulk flow on distance scales exceeding that of the SSC. Our starting
points for the analysis were the findings of \cite{colin} who report a
tentative observation of a backside infall behind the SSC as well as
the Dark Flow reported by \cite{kashlinsky10}.

The paper is organized as follows: The \snf dataset is presented in
Sec.\,\ref{data}, along with the procedure used to combine it with the
Union2 SN compilation. In Sec.\,\ref{anal} the methods used to analyze
bulk flows are described. The data are divided into redshift shells
and fit for a bulk velocity (the \emph{dipole fit} (DF) method based
on \citealt{bonvin}). An alternative approach to test for
anisotropies, the \emph{smoothed residuals} (SR) method \citep{colin}
is employed on the data as well. The results of this study are
presented in Sec.\,\ref{resu}. An analysis of a simplified model of a
SSC-like overdensity is presented in Sec.\,\ref{disc}. Finally, a
discussion of the findings is given in Sec.\,\ref{conc}.

\section{Datasets \label{data}}
The datasets used in this study are a subset of 117 SNe~Ia from the
Nearby Supernova Factory \citep{aldering} dataset described below, and
the Union2 compilation of 557 SNe~Ia \citep{amanullah} as well as the
combination of the two. In order to study the redshift dependence of a
possible anisotropy, the data were divided in redshift shells,
following \citet{colin}. The available number of distances per
redshift bin is summarized in Table\,\ref{aniso_tab}.

\subsection{Nearby Supernova Factory data}

\noindent The \snf is devoted to the study of SNe~Ia in the nearby
Hubble flow ($0.03<z<0.08$) for use in cosmological analyses. It
features a custom-built two-channel Supernova Integral Field
Spectrograph (SNIFS) mounted on the University of Hawaii 2.2 m
telescope at Mauna Kea. SNIFS has a field of view of $6''\times6''$.
It produces a spectrum for each element of a $15\times15$ microlens
array, resulting in a 3D data cube at a given epoch. Each cube
contains the full information of the SN signal, host galaxy and sky as
a function of wavelength $\lambda$ and sky position $x,y$, resulting
in 225 spectra per data cube. The measurement is thus independent of
any filter characteristics and K-corrections. The SN signal is
extracted from the data cube using point spread function fitting
techniques, a calibration procedure that fits for nightly atmospheric
extinction \citep{Buton:2012cr} as well as host galaxy subtraction
\citep{ddt}. A more complete description of \snf, SNIFS and its
operation can be found in \cite{aldering} while more information on
the data processing is presented in
\cite{aldering06} and updated in \cite{scalzo10}.

Lightcurves were synthesized from the flux-calibrated spectra as
recorded with the SNIFS instrument. For this, box filters were applied
roughly matching B, V, and R (see \cite{bailey} for details). The SN
B-band restframe magnitudes at maximum light, $m_B$, were then
extracted using the SALT2 lightcurve fitting algorithm \citep{guy}.
Hence, the distance modulus, $\mu_B$, for each SN can be obtained
after correcting for an empirical width- and color-luminosity relation
\citep{phillips, tripp}:
\begin{equation}
\label{eq:2}
\mu_B=m_B-M+\alpha\cdot x_1-\beta\cdot c,
\end{equation}
where $m_B$, $x_1$ and $c$ are determined in the lightcurve
fit\footnote{Here $x_1$ is a ``stretch factor'' parameterizing the SN
  lightcurve and $c$ describes the the difference in color between the
  observed SN and a template SN after correcting for Galactic
  extinction \citep{guy}.} while $\alpha$, $\beta$ and the absolute
magnitude $M$ are parameters in the fit of the SN Hubble
diagram \citep{guy}.

The dataset used in the analyses is obtained from the full \snf data
set available at the start of this study through application of
quality criteria:
\begin{enumerate}[a)]
\item the SALT2 lightcurve fit includes more than 5 independent epochs,
\item the normalized median absolute deviation (nMAD) of the fit
  residuals over all used filters is smaller than 0.2 mag,
\item there is less than a 20\,\% rejection rate of data points in
  SALT2
\item removal of suspected super-Chandrasekhar SNe. \citep{scalzo12}
\end{enumerate}

A sample of 117 SNe~Ia passed the selection criteria. The lightcurve
parameters and Hubble residuals for most of the SNe have been
published in previous \snf papers \citep{bailey,chotard11}. The
systemic host redshifts are published in \cite{childress13a}. The data
were then grouped in redshift bins as given in Table~\ref{aniso_tab}.
Fig.~\ref{sn_sky} shows the SN distribution on the sky for three
different bins for the Union2 and \snf datasets respectively. The \snf
dataset is limited to declinations easily observable from Mauna Kea
($-25^{\circ}<\delta<65^{\circ}$) and therefore covers approximately
70\% of the sky.

\subsection{Union2 and the Combined Dataset}
\label{sec:combined-dataset}

Additionally we analyzed the Union2 compilation \citep{amanullah} of
577 SNe of which 165 are at redshifts below $z=0.1$
\citep{hamuy96,riess98,perlmutter99,riess99,knop03,tonry03,barris04,krisciunas05,astier06,jha06,miknaitis07,riess07,amanullah08,holtzman08,kowalski08,hicken09}.
However, 109 of these low-redshift SNe are at $z<0.035$
while the \snf dataset has a more even distribution of redshifts.

We checked both datasets for SNe that are close to galaxy clusters
using the NASA/IPAC Extragalactic Database (NED). As the velocities of
such SNe are dominated by virial motion, the redshift of the cluster
should be used instead of that of the SN. We found 11 SNe having a
projected separation within 1~Mpc of a confirmed cluster and with
redshift differences within the velocity dispersion of that cluster.

After analyzing the datasets separately, we combined them to create a
new sample of 279 SNe\footnote{SN~2005eu~\citep{hicken09} was also
  observed by the \snf as SNF20051003-004. We included it in both data
  sets when analysing them separately. Since the results agree we only
  used the Union2 value in the combined dataset.} spanning $0.015 < z
< 0.1$. As the absolute magnitudes, $M$, of the SNe and the parameters
$\alpha$ and $\beta$ of Eq.~\ref{eq:2} were determined separately for
the datasets following the Union procedure, their normalizations may
differ. This would lead to a larger than usual scatter of the Hubble
residuals. Therefore we determined an offset between the distance
moduli in a $\chi^{2}$ fit of a flat $\Lambda$CDM cosmology, with
$\Omega_{M}$ and the absolute corrected magnitude $M$ of SNe the
parameters of the fit. These parameters were determined using all 693
SNe of both datasets and were left blinded to preserve the
impartiality of ongoing \snf cosmology analyses.

\section{Analysis \label{anal}}
For the study of anisotropies and bulk flows present in SN~Ia data two
methodologically distinct techniques were chosen. A \emph{dipole fit}
(DF) method based on \cite{bonvin} is used to determine the bulk flow
velocity in redshift shells. As a cross-check we reimplement the
\emph{smoothed residuals} (SR) method previously used by \cite{colin}. The
significance of both methods is determined by randomly resampling the
directions of the SNe.

\subsection{Peculiar Velocity Dipole \label{sec_dip}}
\noindent A bulk flow toward a certain direction will be observable in
the supernova data as a dipole in the peculiar velocity field. 
In order to quantify the impact of a bulk flow imprinted in the
velocity field of SN~Ia data, luminosity distances
$d_L(z,v_{\mathrm{DF}})$ are fit with a simple dipole model. Following
\cite{bonvin}
\begin{equation}
  \label{v_dip}
  d_{L}(z,v_{\mathrm{DF}},\theta) =d_{L}^{(0)}(z)+d_{L}^{({\rm dipole})}(z,v_{\mathrm{DF}},\theta),
\label{eq:d1_d2}
\end{equation}
where 
\begin{equation}
  \label{d_L_lin}
  d_{L}^{(0)}(z)=c(1+z)\int\limits_{0}^{z}\frac{\text{d}z'}{H(z')},
\end{equation}
$z$ is the cosmological redshift, $v_{\mathrm{DF}}$ is the
dipole velocity amplitude, $\theta$ is the angle between the line of
sight w.r.t.~a single SN and the dipole direction and $H(z)$
represents the Hubble parameter.

The dipole term $d_{L}^{({\rm dipole})}(z,v_{\mathrm{DF}},\theta)$ can be written as
\begin{equation}
  \label{vlin}
  d_{L}^{({\rm dipole})}(z,v_{\mathrm{DF}},\theta)=\frac{v_{\mathrm{DF}}(1+z)^{2}}{H(z)}\cdot\cos(\theta).
\end{equation}
Thus the bulk flow can be determined by comparing the measured
distance moduli, $\mu_{i}$, and measured redshifts, $z_{i}$, to the
prediction from the $\Lambda$CDM Hubble law, i.e.\ by minimizing the
expression
\begin{equation}
  \label{chi2_dip}
\chi^{2}=\sum_i\frac{\left|\mu_i-5\log_{10}\left(\left(d_{L}^{(0)}(z_i)-d_{L}^{({\rm
          dipole})}(z_i,v_{\mathrm{DF}},\theta_{i})\right)/10~\text{pc}\right)\right|^{2}}{\sigma_{\i}^{2}}
\end{equation}
with the bulk flow velocity $v_{\mathrm{DF}}$ and its direction as
parameters. The uncertainties, $\sigma_{i}$, include a
sample-independent intrinsic scatter of $\sim0.15$. Note that because
$d_{L}^{({\rm dipole})}$ is linear in $v_{\mathrm{DF}}$, negative
values of $v_{\mathrm{DF}}$ are permitted mathematically. Fitting
$\chi^{2}$ in opposite directions would result in a sign change of the
velocity. To break this degeneracy we forced $v_{\mathrm{DF}}$ to be
non-negative. Therefore the resulting $\chi^{2}$ value is equal to the
value for the $\Lambda$CDM Hubble law for all directions on one half
of the sky. The uncertainties (68\% CL) of $v_{\mathrm{DF}}$ and its
direction are determined from minimizing $\chi^{2}$, and finding the
parameter values for which $\chi^{2}$ is increased by $\Delta
\chi^2=1$. To compute the significance of a small $\chi^{2}$, the SN
Ia data are resampled 5000 times, i.e.~taking the same $\mu_i$ but
with SN coordinates $(\theta_i,\phi_i)$ randomly interchanged. The
fraction of random realizations that have a smaller $\chi^{2}$ than
the observed one gives us the $p$-value of the measured bulk flow. In
this procedure, we do not account for cosmic variance when determining
the $p$-value and may hence slightly underestimate the uncertainties.
Assuming cosmic variance terms of $\sim$100~km~s$^{-1}$, we expect the
intrinsic dispersion to dominate the error budget for all but the
nearest redshift shell.

In Eq.~\ref{chi2_dip} we do not include the correlation in peculiar
velocity between nearby SNe expected due to large scale structure
\citep{Hui:2005nm,Gordon:2007zw,Gordon:2007sk,Davis:2010jq}, as one
goal of the analysis is to establish the large scale correlation
empirically. The bulk flows tested would result in correlations
between data points that mostly exceed the ones expected on average
due to large scale structure, hence including both would lead to an
overestimation of the covariance. To test this we have compared the
covariance expected for a bulk flow of 300~km~s$^{-1}$ to that derived
from a full power spectrum according to \cite{Hui:2005nm}. The
correlation for a bulk flow is smaller than the full correlation up to
SN separations of $\sim70$~Mpc. As these few pairs of SNe are mostly
at low redshift, we expect no significant effect on the
higher-redshift bins to which we add the most new data.

Here we have restricted the analysis to a dipole in the peculiar
velocities. However, a bulk flow due to a massive attractor like the
SSC is expected to a have a shear component as well which would be
observed in higher multipole orders. An analysis accounting for the
quadrupole and octupole as in \cite{macaulay11} will be presented in a
subsequent paper~\citep{feindt13}.

\subsection{Smoothed Residuals (SR) \label{sr}}
As an alternative means of identifying potential bulk flow motions,
the direction on the sky with the highest deviation of the data from
the isotropic $\Lambda$CDM cosmological model is determined. The
corresponding test statistic is constructed from the error-weighted
Hubble residuals, $r_i$, from Eq.~\ref{chi2_dip}, i.e.
\begin{equation}
  \label{residuals}
  r_{i}=\frac{\mu_i-5\log_{10}\left(d_{L}^{(0)}(z_i)/10\text{pc}
      \right)}{\sigma_{\mu}}.
\end{equation} 
The method of \emph{smoothed residuals} (SR) as described in detail in
\cite{colin} was implemented and further extended for this work. It
can be briefly summarized as follows:\\
The statistic used for such an analysis is
\begin{equation}
\label{smooth_stat}
Q(l_{\rm SR},b_{\rm SR})=\frac{\sum_ir_i\cdot W(l_{\rm SR},b_{\rm SR},l_{i},b_{i})}{\sum_iW(l_{\rm SR},b_{\rm SR},l_i,b_i)},
\end{equation}
where the window function $W(l_{\rm SR},b_{\rm SR},l_i,b_i)\equiv W_i$ reads
\begin{equation}
\label{weight}
W(l_{\rm SR},b_{\rm SR},l_i,b_i)=(2\pi\delta^2)^{-1/2}\exp\left(-\frac{L(l_{\rm SR},b_{\rm SR},l_i,b_i)^2}{2\delta^2}\right).
\end{equation}
Here $\delta$ is a smoothing parameter, which in the following is
chosen as $\pi/2$ in order to identify dipoles in the Hubble
expansion. $(l_{i},b_{i})$ are the coordinates of a SN and $(l_{\rm
  SR},b_{\rm SR})$ those of the direction in which $Q$ is evaluated.
$L$ represents the angular distance between two points.

\citeauthor{colin} used $Q(l_{\rm SR},b_{\rm SR})=\sum_ir_i\cdot
W(l_{\rm SR},b_{\rm SR},l_i,b_i)$ whereas in Eq.~\ref{smooth_stat} a
weighting is introduced that is new to this paper. The weighting
factor $[\sum_iW(l_{SR},b_{SR},l_i,b_i)]^{-1}$ avoids oversampling of
inhomogeneities due to uneven sky coverage, which can lead to
artificial bulk flow signals.

The sky populated with any SN~Ia dataset is scanned in steps
of $10^{\circ}$ in Galactic coordinates $(l_{\rm SR},b_{\rm SR})$. For
every direction the value of $Q(l_{\rm SR},b_{\rm SR})$ is computed
according to Eq.~\ref{smooth_stat} and the maximum and minimum values,
i.e.~$Q_{max}$ and $Q_{min}$, and corresponding directions on the sky
are recorded. To quantify the deviation from the model prediction the
test statistic
\begin{equation}
\label{colineq}
\Delta Q=Q_{max}-Q_{min} 
\end{equation}
is evaluated. The significance of large $\Delta Q$ is determined
by randomly resampling the directions 5000 times; the
$p$-value is the fraction of realizations that have a $\Delta Q$
larger than the measured value. The direction is inferred from
\begin{equation}
\min\left(p_{max},p_{min}\right)
\end{equation}
with $p_{max}$ and $p_{min}$ being the $p$-values with respect to the
$Q_{max}$ and $Q_{min}$ distributions, respectively. For example, in
the $Q_{min}$ case the $p$-value is of course the fraction of MC
measurements that are smaller than the measured $Q_{min}$.

 \begin{figure*}[h]
    \centering
    \includegraphics[width=7.2in]{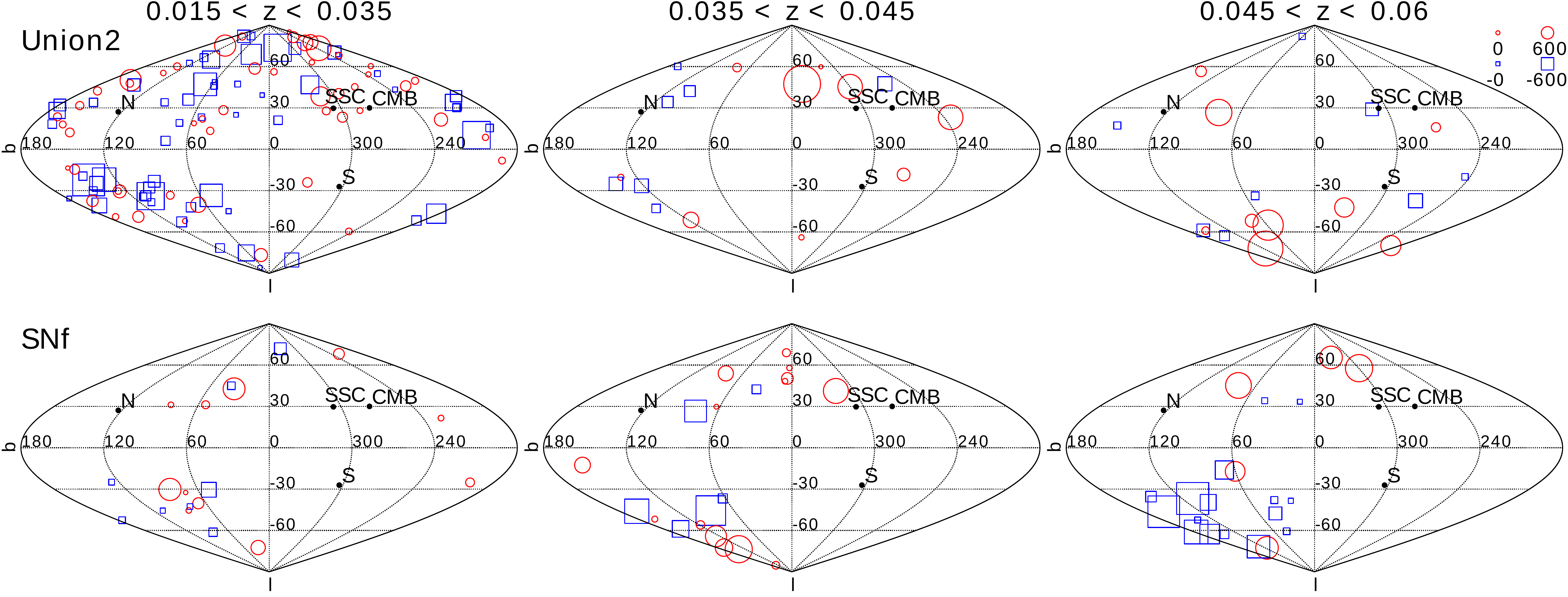}
    \caption{Peculiar velocities of individual SNe determined
        from their distance moduli, $\mu_{i}$, by solving
        Eq.~\ref{eq:d1_d2} for $v_{\rm DF}$. The plots show the
      Union2 (top row) and \snf (bottom row) datasets in the redshift
      bins $0.015<z<0.035$ (left column), $0.035<z<0.045$ (middle
      column) and $0.045<z<0.06$ (right column). The marker
      diameter of each SN is proportional to the absolute
        value of the velocity plus an offset (see the scale at the top
        right), with red circles corresponding to positive velocities
        and blue squares corresponding to negative ones. For
      reference, the directions of the CMB dipole and the Shapley
      Supercluster (SSC) are shown.}
          \label{sn_sky}
    \end{figure*}
 \begin{figure*}[h]
   \centering
    \includegraphics[width=7.2in]{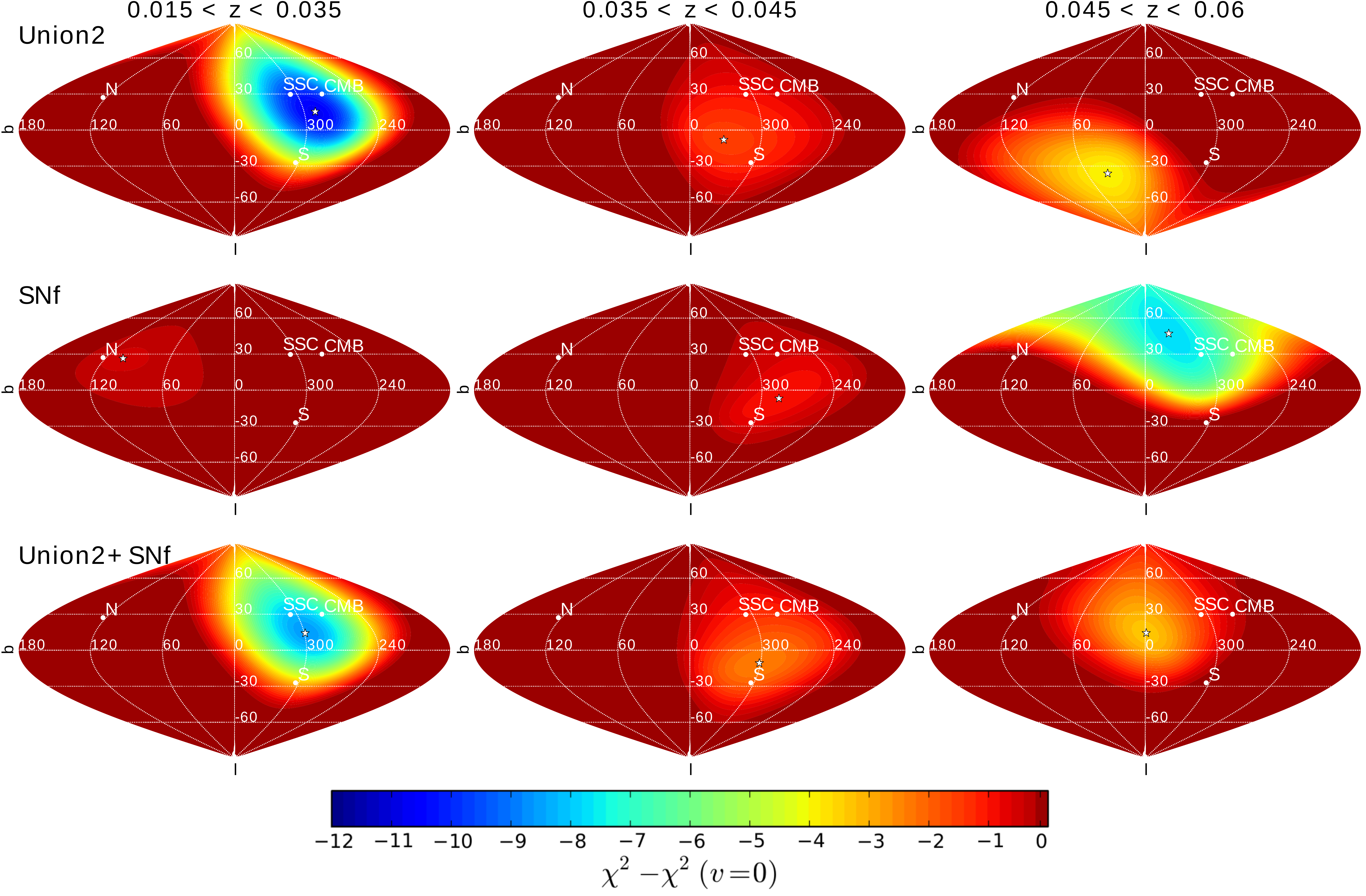}
    \caption{Variation of $\chi^{2}$ for a dipole fit of SNe~Ia from
      the Union2 (top row) and \snf (middle row) datasets and their
      combination (bottom row) as a function of Galactic coordinates
      $(l,b)$ in the redshift range $0.015<z<0.035$ (left column),
      $0.035<z<0.045$ (middle column) and $0.045<z<0.06$ (right
      column). The best fit direction is marked by a star at minimum
      $\chi^{2}$.}
          \label{q_sky}
    \end{figure*}
 \begin{figure*}[h]
    \includegraphics[width=7.2in]{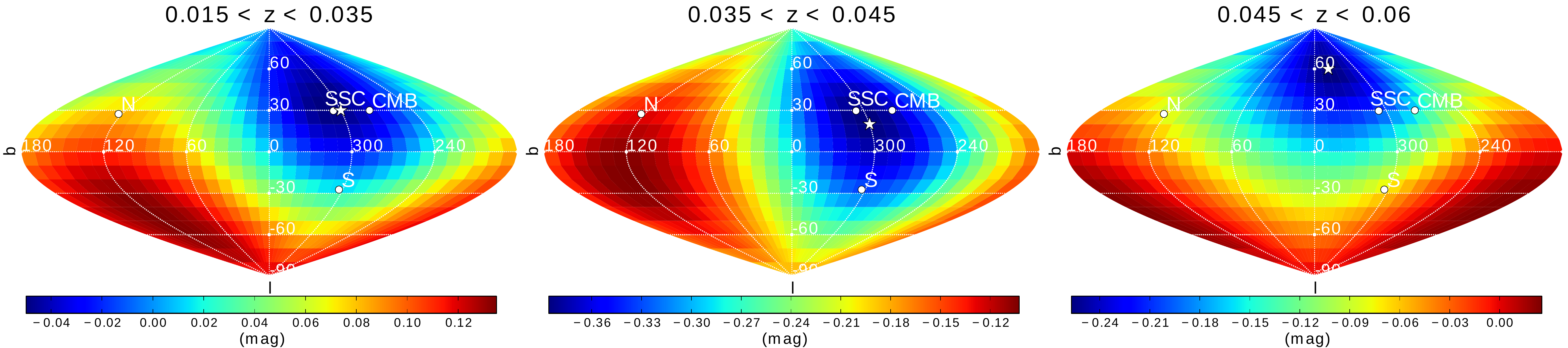}     
    \caption{Magnitude residuals of SNe~Ia from the combined Union2
      and \snf dataset as a function of galactic coordinates $(l,b)$
      after smoothing with a Gaussian window function of width
      $\delta=\frac{\pi}{2}$ in the redshift range $0.015<z<0.035$
      (left), $0.035<z<0.045$ (middle) and $0.045<z<0.06$ (right). The
      bulk flow direction is marked by a star.}
          \label{dip_sky}
    \end{figure*}
    This algorithm was applied to subsamples of the datasets as
    described in Sec.\,\ref{data} in order to study the
    run of potential anisotropies with redshift.

\section{Results \label{resu}}
Table \ref{aniso_tab} shows the velocity and direction of the minimum
variance bulk motion, as reconstructed with the dipole fit (DF), as
well as the amplitude and direction as obtained from the smoothed
residuals (SR) method for each redshift shell for Union2, \snf and the
combined dataset. Also shown are the corresponding $p$-values, i.e.
the chance probability to observe a signal of similar or larger
significance in a homogeneous and isotropic Universe. In the
following, the results of the two analysis methods are presented in
more detail.

\begin{table*}[t]
  \caption[]{Reconstructed directions (in Galactic coordinates) and
    $p$-values of maximum anisotropy according to a dipole fit and the methods of
    smoothed residuals in different redshift bins
    for the \emph{Union2} and \snf datasets and their combination.}
  \centering
      \begin{tabular}{ccc|rrrr|rrr|rrrr}
        \hline \hline &&&&&&&& &\\[-0.2cm]
        \mbox{ Union2} &&&  \multicolumn{4}{c|}{\mbox{
            Dipole fit}} &\multicolumn{3}{c|}{\mbox{ Smoothed residuals}}
        &\multicolumn{4}{l}{\mbox{ Sample characteristics}}\\

        \hline
        &&&&&&&& &\\
        &$N_{\mathrm{SN}}$ &z&
        \multicolumn{1}{c}{$l$}&\multicolumn{1}{c}{$b$}   &
        \multicolumn{1}{c}{$v_{\mathrm{DF}}\,\mathrm{[km~s^{-1}]}$} &
        \multicolumn{1}{c|}{$p$-value}&
        \multicolumn{1}{c}{$l$}& \multicolumn{1}{c}{$b$} &
        \multicolumn{1}{c|}{$p$-value} &
        \multicolumn{1}{c}{$|\vec{\bar{n}}|$} &\multicolumn{1}{c}{$l$}&
        \multicolumn{1}{r}{$b$} &\\[0.1cm]
        \hline 
        &&&&&&&& &\\[-0.2cm] 
        & 109 & 0.015-0.035 & $290(22)$ & $15(18)$ & $292(96)$ & 0.004
        & 300 & 30 & 0.030 & 0.287 & $144$ & $54$    \\
        & 16 & 0.035-0.045 & $331(59)$ & $-7(37)$ & $496(468)$ & 0.316
        & 310 & 20 & 0.230 & 0.200 & $148$ & $17$  \\
        & 17 & 0.045-0.060 & $39(45)$ & $-36(28)$ & $870(490)$ & 0.068
        & 60 & $-30$ & 0.234 & 0.327 & $243$ & $-58$   \\
        & 23 & 0.060-0.100 & $54(93)$ & $-10(53)$ & $509(601)$ & 0.465
        & 70 & $-30$ & 0.724 & 0.340 & $139$ & $-70$   \\[0.1cm] 
        \hline 
        &&&&&&&& &\\[-0.2cm]       
        & 55 & 0.100-0.200 & $256(90)$ & $-18(34)$ & $1238(1975)$ & 0.135
        & 240 & 40 & 0.524 & 0.666 & $120$ & $-64$   \\
        & 62 & 0.200-0.300 & $14(131)$ & $11(75)$ & $1221(1457)$ & 0.327
        & 0 & 60 & 0.644 & 0.805 & $91$ & $-57$   \\
        & 62 & 0.300-0.400 & $257(84)$ & $-36(27)$ & $2590(2841)$ & 0.102
        & 130 & $-80$ & 0.184 & 0.581 & $131$ & $-56$  \\
        & 58 & 0.400-0.500 & $161(48)$ & $28(29)$ & $4190(4014)$ & 0.208 
        & 220 & 60 & 0.337 & 0.211 & $166$ & $12$   \\
        & 44 & 0.500-0.600 & $15(100)$ & $-17(33)$ & $3977(4113)$ & 0.307
        & 20 & 10 & 0.464 & 0.114 & $167$ & $-42$   \\
        & 50 & 0.600-0.800 & $343(81)$ & $-50(43)$ & $5576(4279)$ & 0.107
        & 30 & $-70$ & 0.072 & 0.137 & $127$ & $-7$  \\
        & 60 & 0.800-1.400 & $75(55)$ & $-14(28)$ & $7238(8933)$ & 0.538
        & 280 & $-60$ & 0.949 & 0.364 & $131$ & $37$   \\[0.1cm]
        \hline 
        \mbox{\snf} \mdseries &&& &&&&&&&\\
        \hline 
        &&&&&&&& &\\[-0.2cm] 
        & 20 & 0.015-0.035 & $104(95)$ & $26(44)$ & $229(410)$ & 0.806
        & 330 & 50 & 0.885 & 0.409 & $92$ & $-25$  \\
        & 20 & 0.035-0.045 & $286(70)$ & $-7(42)$ & $484(516)$ & 0.313
        & 270 & 20 & 0.590 & 0.311 & $82$ & $-15$  \\
        & 21 & 0.045-0.060 & $330(48)$ & $48(46)$ & $941(390)$ & 0.006
        & 310 & 60 & 0.004 & 0.497 & $79$ & $-43$  \\
        & 54 & 0.060-0.100 & $252(134)$ & $5(75)$ & $232(360)$ & 0.791  
        & 270 & 30 & 0.941 & 0.394 & $85$ & $-12$  \\[0.1cm] 
        \hline
        \mbox{Union2+\snf} \mdseries &&& &&&&&&&\\
        \hline 
        &&&&&&&& &\\[-0.2cm] 
        & 128* & 0.015-0.035 & $298(25)$ & $15(20)$ & $243(88)$ & 0.010
        & 300 & 30 & 0.074  & 0.252 & $130$ & $45$ \\
        & 36 & 0.035-0.045 & $302(48)$ & $-12(26)$ & $452(314)$ & 0.131
        & 300 & 20 & 0.186  & 0.211 & $105$ & $-4$ \\
        & 38 & 0.045-0.060 & $359(32)$ & $14(27)$ & $650(398)$ & 0.075
        & 340 & 60 & 0.132  & 0.340 & $87$ & $-67$  \\
        & 77 & 0.060-0.100 & $285(234)$ & $-23(112)$ & $105(401)$ & 0.885
        & 0 & $-90$ & 0.999   & 0.324 & $91$ & $-30$  \\[0.1cm] 
        \hline\\
      \end{tabular}
      \begin{flushleft}
        \textbf{Notes.} The rightmost column shows the weighted mean direction
        of SNe, as described in section \ref{sec:probing-systematics}. $^{(*)}$
        SN~2005eu (=SNF20051003-004) is included in both datasets. The Union2
        measurement was used for the combined datasets.
      \end{flushleft}
  \label{aniso_tab}
\end{table*}   

\subsection{Dipole Fit}

We first focus on the nearest redshift shell, $0.015<z<0.035$ ($45
h^{-1}~{\rm Mpc} - 105 h^{-1}~{\rm Mpc} $). The \snf targeted higher
redshifts than this, thus the constraints here are dominated by data
from the Union2 SN compilation. As presented in
Table\,\ref{aniso_tab}, and shown in Fig. \ref{q_sky} there is a flow
observed ($p=0.004$) with a velocity of
$v_{\mathrm{DF}}=292\pm96\,\mathrm{km~s^{-1}}$ towards
$l=290^{\circ}\pm 22^{\circ}$, $b=15^{\circ}\pm 18^{\circ}$. The
combined data show a slightly higher $p$-value of 0.010 than the
Union2 dataset alone for $0.015<z<0.035$. While the direction of the
bulk flow barely changes, the velocity is slightly lower at
$v_{\mathrm{DF}}=243\pm88\,\mathrm{km~s^{-1}}$ for the combined
dataset.

However, as there is some tension between the two samples, we need to
investigate the consistency of the measured bulk flows to assess
whether the samples should be combined in the first place. For this we
adopt a scheme similar to the consistency test described by
\cite{watkins} by calculating the following $\chi^2$ statistics
\begin{equation}
\label{eq:chi2-test}
\Delta^2 = \sum\limits_{i,j}\left(\Delta v_{\mathrm{DF}}^{(i)}\right) 
\left(C^{-1}\right)_{ij} \left(\Delta v_{\mathrm{DF}}^{(j)}\right)
\end{equation}
where $\Delta v_\mathrm{DF}^{(i)}$ are the Cartesian components of the
vectorial difference between the bulk flow estimates from each sample
and $C$ is their combined covariance. For this redshift shell we find
$\Delta^{2}=2.62$ which corresponds to $p$-value of 0.455 for a
$\chi^{2}$ distribution with 3 degrees of freedom. This verifies the
observed tension is not significant.

The direction from the combined sample is compatible both with the SSC
($l=306.44^{\circ}$, $b=29.71^{\circ}$) and the CMB dipole
($l=276^{\circ}\pm 3^{\circ}$, $b=30^{\circ}\pm 3^{\circ}$;
\citeauthor{kogut}, \citeyear{kogut}). Previous studies have already
shown good agreement with the CMB dipole: e.g.\,\cite{colin}
found a bulk flow of
$v_{\mathrm{bulk}}=250^{+190}_{-160}\,\mathrm{km~s^{-1}}$ towards
$l=287^{\circ}$, $b=21^{\circ}$ at a radius of $100\,h^{-1}$~Mpc using
a maximum likelihood approach. \cite{watkins} computed
$v_{\mathrm{bulk}}=416\pm78\,\mathrm{km~s^{-1}}$ in the direction of
$l=282^{\circ}$, $b=60^{\circ}$ at the same scale. Another general
confirmation of these results comes from the study in \cite{lavaux}
who report $v_{\mathrm{bulk}}=473\pm128\,\mathrm{km~s^{-1}}$,
$l=220^{\circ}$, $b=25^{\circ}$ for the $100\,h^{-1}$ Mpc frame.
Furthermore, \cite{macaulay} find a dipole velocity of
$v=380^{+99}_{-132}\,\mathrm{km~s^{-1}}$, $l =
295^{\circ}\pm18^{\circ}, b = 14^{\circ} \pm 18^{\circ}$ and
\cite{nusser} find a bulk flow of $257 \pm 44\,\mathrm{km~s^{-1}}$
toward $l=276^{\circ} \pm 6^{\circ}, b=10^{\circ}\pm 6^{\circ}$. On
the other hand, \cite{courteau} have reported that this shell is at
rest with respect to the CMB, finding the bulk flow to be only
$70^{+100}_{-70}\,\mathrm{km~s^{-1}}$. Our result is consistent at
$1.2\sigma$ with this amplitude. Hence the results for nearby flow
velocities obtained here are compatible with the literature. Of course
the SNe in this shell are common to several of the studies.

We now turn to the next further redshift shell, $0.035<z<0.045$, where
the data from \snf starts to dominate the available SN statistics.
While the individual datasets show no significant evidence for a bulk
motion, the combined datasets lead to a $p$-value of 0.131 for the
observed anisotropy to be a random coincidence. While not very
significant, it is notable that the best-fit direction is again well
aligned with the SSC direction, strengthening the evidence that the
observed bulk motion of the shell is real. The upper boundary of the
redshift shell was chosen by \cite{colin} such that it
intersects the center of the SSC\footnote{The SSC roughly extends from
  $0.035<z<0.055$ with its center located at $z\sim 0.046$.}. Hence,
if the SSC is responsible for the motion, one would expect the flow to
retain its direction.

The redshift bin ($0.045<z<0.06$) intersects the SSC at the lower
redshift boundary. Hence, if the the missing component of the LG
velocity is due to the infall into the SSC, one would expect the bulk
motion to reverse its direction in the hemisphere around the SSC while
it remains the same in the other hemisphere. For the Union2 data,
($0.045<z<0.06$) the direction of anisotropy appears to be reversed to
$l=39^{\circ}\pm 45^{\circ}$, $b=-36^{\circ}\pm 28^{\circ}$. This is
in agreement with the turnover seen by \cite{colin} using the same
data. However, with a $p$-value of $p=0.068$ the reversal is not very
significant.

The \snf data do not support this reversal. Rather, the bulk flow in
the corresponding shell maintains its global direction with
$l=330^{\circ}\pm 48^{\circ}$, $b=48^{\circ}\pm 46^{\circ}$ with a
rather small $p$-value of 0.006. The best fit bulk flow velocity
$v_{\mathrm{DF}}=941\pm390\ \mathrm{km~s^{-1}}$ appears to be large, but
due to its large statistical errors not inconsistent with the bulk
flow motions obtained at lower redshifts. When combining the \snf and
Union2 datasets, this bulk flow beyond the SSC is not as significant
as when using \snf data alone, but still no turnover can be observed:
In the shell at $0.045<z<0.06$ the bulk flow is shifted slightly away
from this, pointing toward $l=359^{\circ}\pm 32^{\circ}$,
$b=14^{\circ}\pm 27^{\circ}$ with a lower velocity of
$v_{\mathrm{DF}}=650\pm398\ \mathrm{km~s^{-1}}$.

As with the redshift shell at $0.015<z<0.035$ we observe tension
between the datasets in this shell. According to
Eq.~\ref{eq:chi2-test} we find significant tension with
$\Delta^{2}=8.59, p=0.035$. We would like to understand the source of
this disagreement between the Union2 and \snf data in the
$0.045<z<0.06$ shell. The underlying distribution of SN peculiar
velocities on the sky from the Hubble fit is depicted in
Fig.\,\ref{sn_sky}. The size of the markers correlates with the
amplitude of the velocity while color and shape correspond to the
direction. When looking at the Union2 data for redshifts
$0.045<z<0.06$ (upper right plot) it becomes clear that the putative
turnover in flow direction is not exclusively induced by SNe falling
into the SSC, as previously claimed by \citet{colin}, but also driven
by two distant SNe in the direction opposite the SSC, SN~1995ac
\citep{riess99} and SN~2003ic\footnote{This SN has been previously
  identified as an outlier in the host-luminosity relation
  \citep{Kelly:2009iy}. Furthermore we determined that it is located
  in the galaxy cluster Abell 0085. The difference in redshifts is
  insignificant ($\Delta z = 0.00007$).} \citep{hicken09}. In this
shell SNe in the vicinity of the SSC should otherwise be significantly
blueshifted giving positive residuals, however Union2 lacks SNe there.
We hence find that the Union2 compilation has insufficient area
coverage in this critical redshift shell to address the question of
whether the SSC is the source of the CMB-dipole. For the \snf sample,
on the other hand, there are two redshifted SNe that show large
residuals ($\sim 1.5 \sigma$) behind the SSC while SNe in the opposite
part of the sky are mostly blueshifted.

We tested the effect of removing either or both of the two Union2 SNe
in question (1995ac and 2003ic). When removing one of them at a time,
we find that the tension decreases to $\Delta^{2}$-values of 6.37
($p=0.095$) and 6.32 ($p=0.097$) respectively. Removing both decreases
the tension to $\Delta^{2}=4.15, p=0.246$. Without SNe 1995ac and
2003ic the amplitude of $v_{\mathrm{DF}}$ determined from Union2 (cf.
Table \ref{aniso_tab}) alone decreases to $498\pm 573$~km~s$^{-1}$
while its direction shifts to $l=35^{\circ}\pm 66^{\circ},
b=-20^{\circ}\pm 28^{\circ}$. The effect of removing both SNe on the
combined dataset is insignificant as we find $v_{\mathrm{DF}}=629\pm
336$~km~s$^{-1}$ towards $l=351^{\circ}\pm 40^{\circ}, b=37^{\circ}\pm
37^{\circ}$. Furthermore we note that the probability of finding a
$p$-value of less than 0.035 for at least one of four measurements is
13.3\% and hence finding this tension in one of four redshift shells
is not as significant as it may seem by itself.

Finally, for the $0.06<z<0.1$ shell, no bulk motion is detected for
any of the samples --- the best fit velocity is consistent with zero
with an associated uncertainty of $\sim400$~km~s$^{-1}$. This
contradicts the results of \cite{kashlinsky10} who measured a bulk
flow of $\sim1000$~km~s$^{-1}$ aligned with the CMB dipole direction
at the same distance. For a better comparison to their results, we fit
the absolute value of the dipole velocity while keeping its direction
fixed at the CMB direction. This yields a dipole velocity of $26\pm
236$~km~s$^{-1}$, i.e we rule out a bulk flow as seen by
\citeauthor{kashlinsky10} at $\sim 4 \sigma$.

\subsection{Smoothed Residuals Analysis}
The results for the smoothed residual analysis can be found in the
second column to the right of Table \ref{aniso_tab}. The
magnitude residuals for the combined dataset for $0.015<z<0.06$ are
shown in Fig.~\ref{dip_sky}. The directions determined by this method
are fully compatible in reconstructed bulk flow directions and
significance levels with what was already found in the dipole fit.
Apparent differences occur in bins with poor statistics where
basically noise is fit and the error bars are correspondingly large.

Comparing the directions determined for the Union2 sample to those of
\cite{colin}, we find agreement within $\sim 15^{\circ}$ for most
shells for which we find $p<0.5$. The scientifically most interesting
shell is that at $0.045<z<0.06$. We find an insignificant deviation of
15$^{\circ}$ in that shell. However, note that there are two
differences between our analysis and that of \cite{colin}. Firstly, we
divide their test statistic $Q$ by the sum of the weights $W_{i}$ to
avoid artificially large values in regions of the sky where more SNe
were observed. Secondly, we set the redshifts of SNe near galaxy
clusters to the cluster redshifts. One of these SNe is SN1993O in the
cluster Abell 3560 ($z=0.04985$). As its redshift provided in the
Union2 compilation is 0.0529\footnote{This redshift differs from the
  one obtained from NED ($z=0.514$ in CMB restframe).}, the SN appears
to be moving toward us instead of away from us when using the redshift
of Abell 3560. This increases the statistic $Q$ for directions
opposite to that of the SSC. However, if we use the redshifts that are
provided in the Union2 compilation, we find a minimum of $Q$ at
$l=50^{\circ},b=-10^{\circ}$, which deviates from the result of
\citeauthor{colin} by 37$^{\circ}$. Therefore our implementation of
the smoothed residuals is indeed not biased by regions of larger
sample density.

Furthermore a larger disagreement between the methods is found at
redshift $0.1<z<0.4$ where SNe from SDSS \citep{holtzman08}, which are
located on a thin stripe on the sky, dominate the dataset.

\subsection{Systematic uncertainties}
\label{sec:probing-systematics}

Systematic errors associated with SNe~Ia as distance indicators have
been scrutinized carefully for the measurement of the expansion
history of the Universe. Some of the main sources of redshift- or
sample-dependent systematic error are connected to flux zero points,
K-corrections and Malmquist bias \citep{amanullah,Regnault:2009kd}.
They can also conspire in non-trivial ways, such as in the case of the
``Hubble bubble'' \citep{Jha:2006fm}, an apparent 6\% increase of the
local Hubble parameter. The Hubble bubble was explained in
\citet{Conley:2007ng} as the result of a combination of a bias in
color, i.e. the closest SNe in the sample show more reddening, and the
choice of the reddening correction parameter $\beta$. Below we discuss
our measurement of bulk flows in the context of (1) a redshift or
sample-dependent systematic error on the distance modulus and (2)
direction-dependent systematic errors.

To first order, a redshift-dependent systematic error
averages out when inferring directional bulk flows. It is only because
of e.g. anisotropic sky coverage, that the averaging process can
result in a residual bias.
For the bulk flow analyses the distribution of SNe on the sky has a
great impact on the sensitivity in a given direction, e.g.\ if most
observations lie in a direction perpendicular to the bulk flow, no
good constraints on its value can be derived. Alternatively if a SN
sample that lies in a preferred direction has a systematic bias, this
will induce a false bulk flow signal.\\
To quantify this effect, we calculate the weighted mean direction
$\vec{\bar{n}}$ of the datasets, i.e.\
\begin{equation}
    \label{eq:3}
    \vec{\bar{n}}=\frac{\sum_{i}w_{i}\vec{n}_{i}}{\sum_{i}w_{i}}\quad
    \mathrm{with}\quad
    w_{i}=\frac{1}{\sigma(\mu_{i})^{2}}\frac{d_{L}^{({\rm
          dipole})}(z_{i},v_{\mathrm{DF}})}{d_{L}^{(0)}(z_{i})}
\end{equation}
where $\vec{n_{i}}$ is the normal vector for a SNe's coordinates. The
weights, $w_{i}$, are calculated from the uncertainty
$\sigma(\mu_{i})$ of the distance modulus and the ratio of
$d_{L}^{({\rm dipole})}$ to $d_{L}^{(0)}$ (see Eq.~\ref{eq:d1_d2}) for
an arbitrary velocity where the latter corresponds to the fact that in
our weighting scheme SNe at higher redshifts have a smaller weight on
the inferred bulk motion. Note that the dependence of $d^{\rm
  (dipole)}$ on $v_{\rm DF}$ cancels when dividing by the sum of the
weights, as it is linear in $v_{\rm DF}$. For a perfectly homogeneous
distribution of observations $\vec{\bar{n}}$ would vanish. In that
case the sources of systematic errors in cosmological analyses would
average out in the measurement of the bulk flow. If, on the other
hand, the amplitude of $\vec{\bar{n}}$ is non-zero, the systematic
effects would shift proportionately the resulting bulk flow in the
direction of $\vec{\bar{n}}$.

It can be expected that the different SN samples, which where obtained
with different telescopes and analysis pipelines, will have a
different degree of bias, i.e. all brighter or dimmer than expected.
\citet{amanullah} has investigated such biases within the Union2
compilation, and at the level of the statistical errors that range
from 0.01 to 0.05 magnitudes, has not found evidence for their
presence. Furthermore the effect of redshift-dependent systematic
uncertainties, e.g. reference star colors, which are the largest
contributor to the systematic uncertainties (see \citealt{amanullah}),
is considerably smaller because the portion of our SN sample with a
statistical weight spans a small redshift range. Assuming that all SNe
in a redshift shell are biased by $\delta m$ magnitudes, the resulting
bias in velocity would be $ \delta \vec{v}~\mathrm{[km~s^{-1}]}
\approx 1.3 \cdot 10^{5} \cdot z \cdot \delta m\cdot \vec{\bar{n}} $.
Hence, for the lowest redshift bin ($|\vec{\bar{n}}|=0.252$), assuming
a bias of 0.03 magnitudes, one would obtain a velocity bias of $\sim
25$ km~s$^{-1}$. For the redshift shell 0.035-0.45, one would obtain a
velocity bias of $\sim$~33~km~s$^{-1}$, while for the redshift bins
0.045-0.06 and 0.06-0.1 one would obtain a velocity bias 70 and
100~km~s$^{-1}$, respectively. This potential bias should be put into
context of the statistical error\footnote{The achievable statistical
  error can be approximated as $ \sigma_v~\mathrm{[km~s^{-1}]} \approx
  3.6 \cdot 10^{4} \cdot z \cdot N_{\rm SN}^{-1/2}$, where $N_{\rm
    SN}$ is the number of SNe. We have assumed an isotropic
  distribution and a distance modulus uncertainty of 0.15.}. For all
shells it can be seen that such a bias would be smaller than the
statistical error. Note also that the bias is a vector quantity and
hence will have the strongest effect on the velocity if it is parallel
or antiparallel to the bulk velocity --- otherwise it will mainly
change the direction.

To absorb any bias that may be present due to a constant
shell-dependent magnitude offset, we added a \emph{monopole term} to
the luminosity distance in Eq.~\ref{v_dip}:
\begin{equation}
  \label{eq:1}
  d_{L}(z,v_{\mathrm{m}},v_{\mathrm{DF}},\theta)
  =d_{L}^{(0)}(z)+d_{L}^{({\rm monopole})}(z,v_{\mathrm{m}})+d_{L}^{({\rm dipole})}(z,v_{\mathrm{DF}},\theta)
\end{equation}
This term corresponds to a constant radial velocity $v_{\rm m}$ in the
redshift shell that is analyzed and $d_{L}^{({\rm monopole})}$ is
calculated exactly as $d_{L}^{({\rm dipole})}$ according to
Eq.~\ref{vlin} but without the dependence on $\cos(\theta)$. For the
SN magnitudes this translates to an equal shift of all luminosities in
that bin, while a dipole term causes SNe in one half of the sky to
appear brighter than expected from their redshift and those on the
other half to appear fainter. We generally find monopole velocities
that are consistent with zero and all effects on the dipole velocities
are within $1\sigma$ of the fit without a monopole. In particular for
the 1st and 4th bin, the best fit velocities have essentially not
changed, indicating the absence of a potential bias due to a magnitude
offset. For the case of the 4th bin, where we fixed the direction to
the CMB dipole in order to compare to \cite{kashlinsky10}, the dipole
velocity changes from $26\pm 236$~km~s$^{-1}$ for a fit without monopole term
to $-32\pm 271$~km~s$^{-1}$ with monopole term.

As SN statistics for redshifts $0.03<z<0.1$ are still low, one needs
to be careful when combining two datasets with $\vec{\bar{n}}$
pointing in opposite directions. For the combination of the Union2 and
\snf datasets, we have therefore checked that this is not the case,
finding that the angles between $\vec{\bar{n}}$ do not exceed $\sim
90^{\circ}$ for all shells (see the rightmost column of
Table~\ref{aniso_tab}). Additionally the amplitudes of $\vec{\bar{n}}$
are smaller than 0.5 for all shells with $z<0.1$ and smaller than 0.35
for the combined data. For higher redshifts the absolute value can be
larger, especially for $0.1<z<0.4$ because the SNe from SDSS
\citep{holtzman08} are clustered in one direction.

We also tested alternative corrections of the SN magnitude
(Eq.~\ref{eq:2}) that we can apply to the \snf dataset. As was shown
in \cite{bailey}, using the ratio of spectral flux at 642~nm and
443~nm reduces the intrinsic scatter further than using the SALT2
light-curve fit parameters $x_{1}$ and $c$. We applied this correction
to 93 of the 117 \snf SNe and compared the results with those for the
SALT2-corrected distance moduli of the same SNe, finding the changes
to be below $1\sigma$ for the direction and below
$0.2\sigma$ for the bulk flow velocity.

Recent studies have shown that the standardized magnitudes
of SNe~Ia are correlated with the properties of their host galaxies,
such that SNe in galaxies with higher stellar mass $M_{*}$ are
brighter on average
\citep{Kelly:2009iy,sullivan10,lampeitl10,childress13b}. To test the
impact of this on our analysis, we implemented a mass step function,
i.e. we split the \snf dataset at
$\log\left(M_{*}/M_{\sun}\right)=10$ (using host mass data presented
in \citealt{childress13a}) and allowed for different normalizations of
the split samples. There is no change in our results for the bulk flow
(less than $0.2\sigma$ for the bulk flow velocity.) Similarly, we
tested extending the SALT2 correction by adding the logarithm of the
host metallicty from \cite{childress13a}. This limited the dataset to
68 SNe. Again our results did not change significantly (by less than
$0.1\sigma$ for both bulk flow direction and velocity).

Finally, we turn to explicitly direction-dependent systematic
errors. As the effect of extinction by dust in the Galaxy is
anisotropic, it is expected to have a larger contribution to the
systematic error of the bulk flow than on cosmological parameters. To
assess the effect of such uncertainties, we increased the distance
modulus of each SN by 10\% of the reported extinction in its direction
\citep{schlegel}. This changes the inferred bulk flow and its
direction by less than $0.05\sigma$. Hence Milky Way extinction can be
considered a small source of systematic uncertainties for these data
sets. Improper atmospheric extinction corrections can also lead to a
directional bias. However, it is usually an integral part of a
calibration procedure that relies on few assumptions, hence resulting
in a small contribution to the total magnitude error (see e.g.\
\citealt{Regnault:2009kd,Buton:2012cr}) that can be safely neglected for
this work.

\section{Searching for Matter Overdensities}
\label{disc}

\noindent Applying the dipole fit (DF) method as well as the method of
smoothing residuals (SR) to the datasets has shown a bulk flow that
continues to point towards the CMB dipole direction and extends beyond
the SSC. A reversal of the bulk flow direction at the distance of the
SSC was rejected (see Table\,\ref{aniso_tab} and Fig.\,\ref{q_sky}
lower panel) in the \snf and combined data, while the apparent
backside infall in Union2 was identified to stem from two SNe located
in the opposite direction of the SSC that are redshifted, instead of
blueshifted. In what follows, we chose an alternative, model-dependent
approach to quantify the constraints of the SN data on an SSC-like
attractor scenario. We thereby test for the presence of a single,
massive object along the line of sight of the SSC. We restrict our
study to the simplified model of a spherical attractor of constant
density $\rho$ with a radius of $R=50$~Mpc, which therefore has a
total mass of
\begin{equation}
  \label{eq:spherical_attractor}
  M_{\mathrm{attractor}}=\frac{4\pi R^{3}}{3}\rho.
\end{equation}
Because the observed bulk motion at larger redshifts is only
marginally significant, this approach is chosen in favor of a more
conventional reconstruction of the entire galaxy density field (as in
e.g.~\citealt{erdogdu2006,lavaux}) which would require larger SN
statistics for meaningful results.

Furthermore, we are interested in measuring the overdensity at the
distance of the SSC and beyond where we contribute the most new data.
At lower redshifts, where the overdensities have already been mapped,
our model will be much less accurate as we do not include known (but
less massive) structures. 

In general, local variations in density, $\rho(\vec{r})$ (where
$\vec{r}$ is a position vector of proper distance originating at the
attractor), will result in additional contribution to the redshift.
Perturbing the distance modulus $\mu=5\log(d_{L}^{(0)})$ (with
$d_{L}^{(0)}$ as per Eq.~\ref{d_L_lin}) by an extra peculiar velocity
induced redshift $z_{\mathrm{p}}$ yields a measured distance
modulus\footnote{For an additional (positive) redshift
  $z_{\mathrm{p}}$ the measured distance modulus at $z$ is actually
  smaller than predicted.}
\begin{equation}
\label{mu_per}
\mu_i=\mu(z_i-z_{\mathrm{p}}).
\end{equation}
The peculiar velocity $v_{\mathrm{p}}$ of a single SN in the
gravitational field of an overdensity
$\delta(\vec{r})=\left(\rho(\vec{r})-\rho_c\Omega_{M}\right)/\rho_c\Omega_{M}$
according to \cite{peebles} and \cite{munoz} is given by
\begin{eqnarray}  
\label{v_p}
\vec{v_{\mathrm{p}}}(\delta,\vec{x})&=&\frac{afH}{4\pi}\int\frac{\vec{y}-\vec{x}}{|\vec{y}-\vec{x}|^{3}}\delta(\vec{y})\mathrm{d}^{3}\vec{y}\nonumber\\
&\approx&\frac{\Omega_{M}^{0.55}H_{0}}{4\pi(1+z)}\int\frac{\vec{y}-\vec{x}}{|\vec{y}-\vec{x}|^{3}}\delta(\vec{y})\mathrm{d}^{3}\vec{y}
\end{eqnarray}
where $\vec{x}$ and $\vec{y}$ are the proper distance vectors pointing
from the center of gravity of the overdensity to a particular SN and
to each mass element of the attractor, respectively, and $\rho_{c}$ is
the critical density of the universe. Projecting
$\vec{v_{\mathrm{p}}}$ along the line of sight the resulting
additional redshift is given by
\begin{equation}  
\label{z_p}
  z_{\mathrm{p}}(\delta,\vec{x})=-\frac{v_{\mathrm{p}}(\delta)}{c}\frac{\vec{x}+\vec{s}}
  {|\vec{x}+\vec{s}|}\cdot\frac{\vec{s}}{|\vec{s}|}
\end{equation} 
where $\vec{s}$ denotes the vector between the observer and a SN.

In this study $\delta(\vec{r})$ is assumed to be constant within a
sphere with radius $R=50$~Mpc corresponding to the size of the SSC,
hence the total mass of the attractor is given by rewriting
Eq.~\ref{eq:spherical_attractor} as
\begin{equation}
  \label{eq:4}
  M_{\mathrm{attractor}}=\frac{4\pi R^{3}}{3}\rho_{c}\Omega_{M}(1+\delta).
\end{equation}
The overdensity profile within the sphere is of little importance to
our analysis as most SNe are outside its boundary. However, as a
single SN inside the sphere can have a large effect on the results, we
introduce weights, $w_{i}$, that are defined as
\begin{equation}
  \label{eq:5}
  w_{i}=\left\{
      \begin{array}{ll}
        \sin^{16}\left(\frac{\pi}{2}\frac{r_{i}}{R}\right)&\textrm{if}\ r_{i}<R\\
        1&\textrm{if}\ r_{i} \geq R
      \end{array}\right. ,
\end{equation}
where $r_{i}$ is the distance from the center of the overdensity. The
overdensity, $\delta(\vec{r})$, is determined by minimizing the
expression
\begin{equation}
  \label{chi2}
  \chi^{2}=\sum_i\frac{\left|\mu_i-\mu\left(z_i-z_{\mathrm{p}}(\delta)\right)\right|^{2}}{\sigma_{\mu}^{2}}w_{i}^{2}  .
\end{equation}
Varying the redshift of such a hypothetical concentration along
the direction $(l_{\rm{SSC}},b_{\rm{SSC}})$ towards the SSC and
minimizing the expression in Eq.~\ref{chi2} yields the corresponding
overdensity at a redshift $z$,
i.e.~$\delta=\delta(z,l_{\rm{SSC}},b_{\rm{SSC}})$, necessary to
account for the peculiar motions (see Eq.~\ref{v_p}) present in the
data.

\begin{figure}[t]
   \includegraphics[width=3.6in]{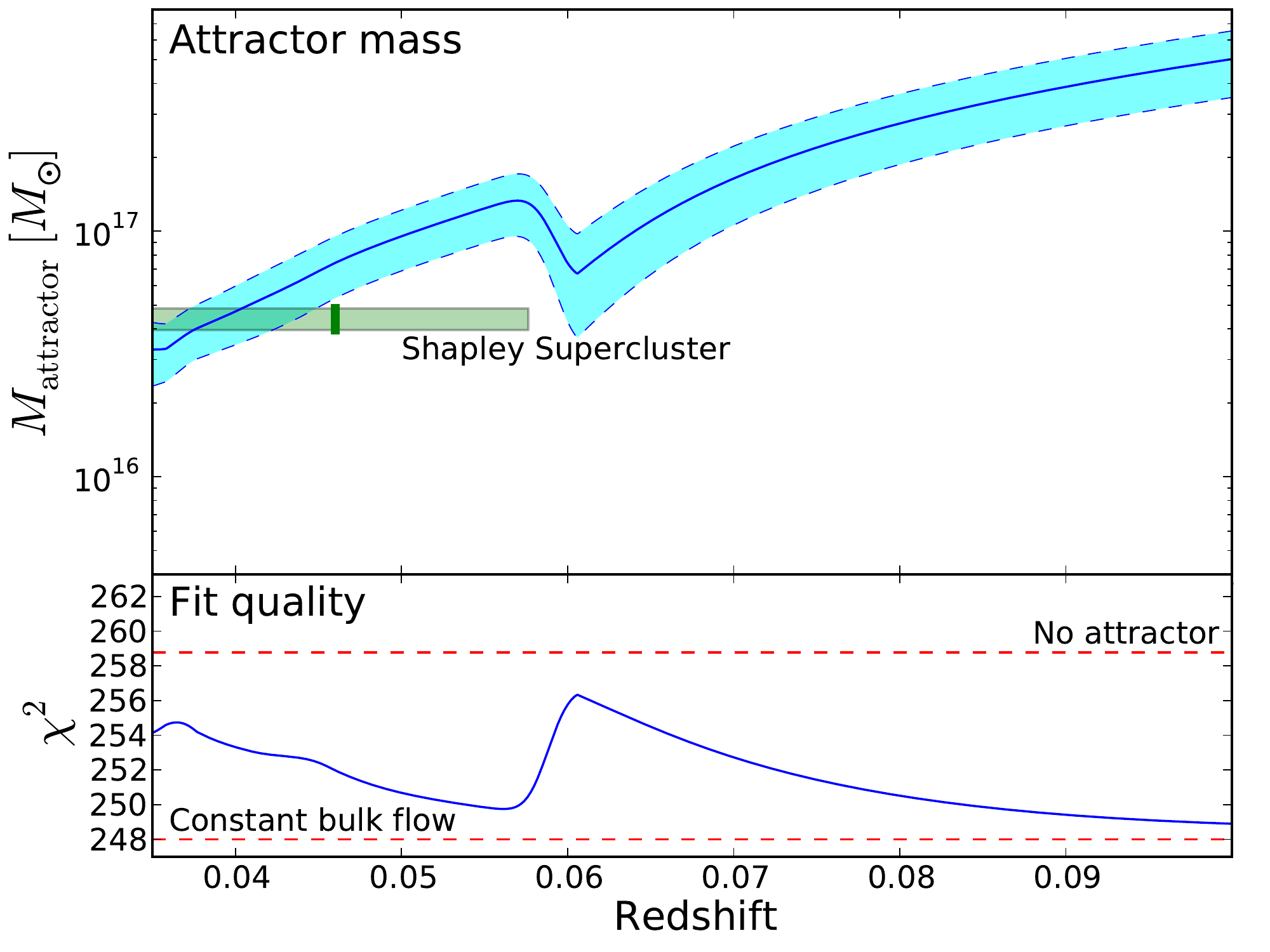}
   \caption{Attractor mass $M_{\textrm{attractor}}$ as a function of
     redshift $z$ accounting for the SNe~Ia peculiar velocities in the
     combination of the Union2 and \snf datasets along with $\chi^2$
     fitting via Eq.~\ref{chi2}. The blue shaded area shows the
       uncertainty of the mass determined for a given
       attractor redshift. The green box shows the mass range as well
     as the approximate size of the SSC from \cite{munoz}. The red
     dotted line shows the $\chi^{2}$ value for a constant bulk flow.}
         \label{ssc}
   \end{figure}

In Fig.~\ref{ssc} the required attractor mass for the peculiar
velocity field extracted from all SNe at $z<0.1$ in the combination of
the Union2 and \snf datasets is shown as a function of redshift $z$ in
the direction of the SSC. As the overdensities at lower redshift have
already been mapped, we restricted the redshift of the model attractor
to $0.035<z<0.1$ Additionally, the corresponding $\chi^2$-values for
the fit with Eq.~\ref{chi2} are shown as a measure of the fit quality.
A mass placed at large distances asymptotically approaches the case of
a constant bulk velocity. This is because an attractor at larger
distance leads to a more homogeneous peculiar velocity field (and thus
less shear) in the nearby Universe. We find that a mass concentration
in the proximity of the SSC is only marginally consistent with our
data. While an attractor at the location of the SSC ($z\sim0.045$) is
disfavored at a level of $2.1~\sigma$ ($\Delta \chi^{2}= 4.4$),
compared to a constant bulk velocity determined by a dipole fit of the
same data,\footnote{This result does not contradict our rejection of a
  Dark Flow in section \ref{resu} because this analysis uses all SNe
  at $z<0.1$ and therefore SNe at low redshifts contribute to it.}
this $\Delta \chi^{2}$ decreases rapidly with redshift and disagrees
by less than 2~$\sigma$ at redshifts $0.046 < z < 0.058$. This
conclusion is based purely on the shape of the gravitational field,
and not the mass itself. The mass that is obtained for the location of
the SSC is larger by a factor of two or more compared to current mass
estimates of the SSC \citep{munoz,2011MNRAS.417.2938S}\footnote{
  \cite{lavaux11} found a higher values for the mass of the SSC
  assuming a larger radius of the SSC.}, suggesting the need for an
additional gravitational source.

\begin{figure}[t]
   \includegraphics[width=3.6in]{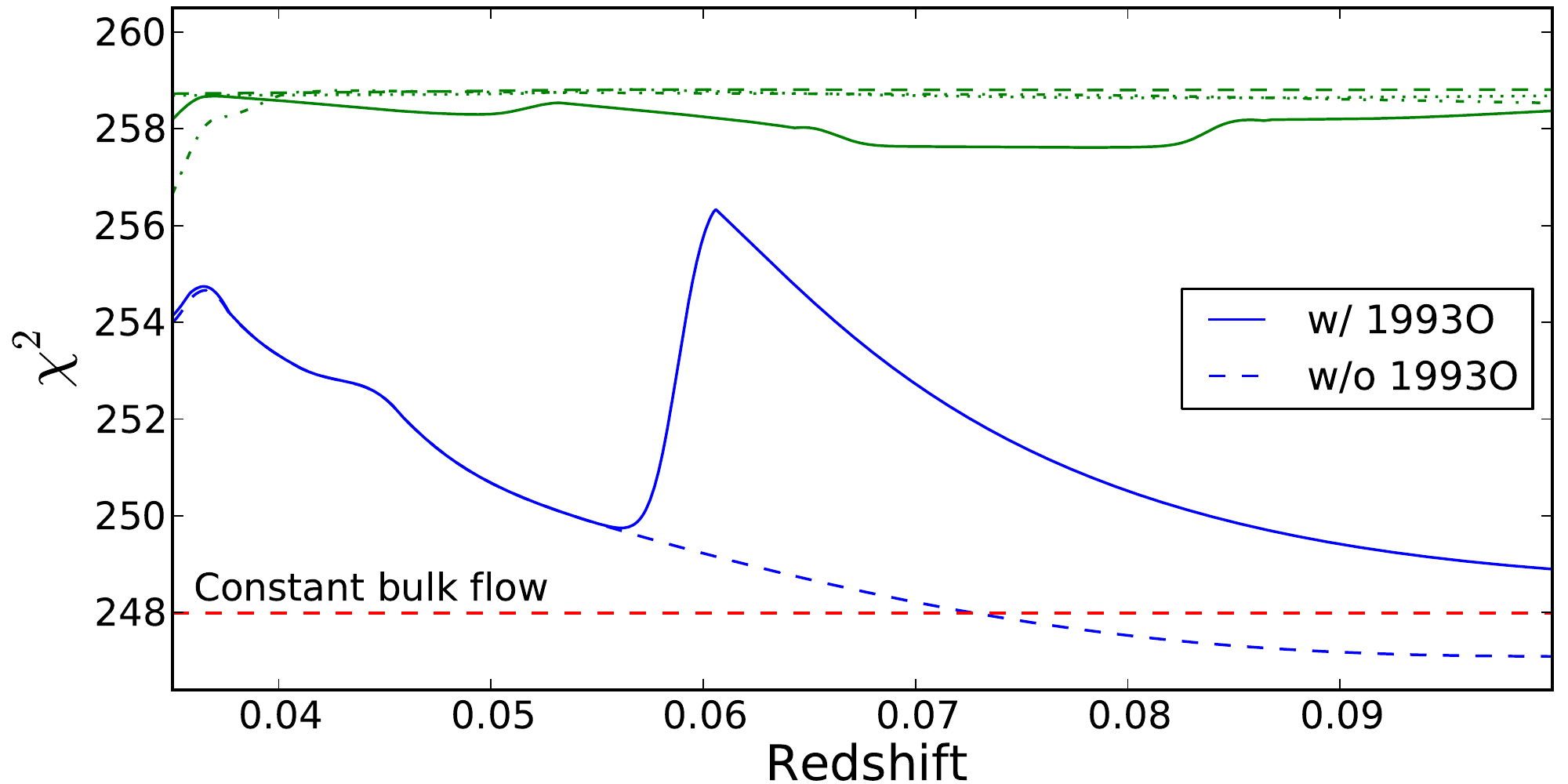}
   \caption{$\chi^{2}$ values for attractor fits in the direction of
     the SSC with and without SN1993O (blue lines) as well as
     for directions 90$^{\circ}$ north (solid green), south
     (dotted green), east (dashed green) and west
     (dotted green) of this in Galactic coordinates.}
         \label{ssc-4-directions}
   \end{figure}

An attractor mass at a redshift further out leads to better fit
results, however. The required overdensity thereby asymptotically
grows with the square of the distance, i.e. $\delta\propto d^{2}$. We
have investigated if the SSC direction is indeed a preferred direction
by repeating the fit placing the attractor in directions that are
perpendicular to the SSC direction. Fig. \ref{ssc-4-directions} shows
the evolution of the $\chi^{2}$ values for these directions that show
no significant decrease for any attractor distance. Hence a search
focused on an attractor in the SSC direction is fully justified.

Furthermore, we tested whether the assumed radius of the attractor
influences the fit results. As the velocity field outside the
overdensity only depends on the total mass of the attractor, we expect
this influence to be small. However, as the velocity field inside is
sensitive to the radius, SNe inside the overdensity can affect the fit
results. Five SNe are within the boundary of the attractor for some
redshifts but their effect is limited by the deweighting according to
Eq.~\ref{eq:5}. Only for SN~1993O the effect remains noticeable as a
dip in the $\chi^{2}$ values and a bump in the attractor mass around
$z\sim0.05$, both of which become narrower for smaller radii. SN~1993O
is located in the galaxy cluster Abell 3560 ($z=0.04985$), 28~Mpc from
the center of the SSC and therefore within the assumed spherical
overdensity. For an attractor radius of 50~Mpc only the high-redshift
edge of the $\chi^{2}$ dip is visible as a large increase in the
$\chi^{2}$ value around at $z\sim 0.06$. This is not only caused by
the SN being inside the boundary of the attractor but rather by its
proximity to the attractor in general. SN~1993O appears to be moving
toward us and therefore an attractor placed behind it will lead to a
much larger residual than an attractor in front of it. This effect is
not seen immediately at the redshift of the SN because it is
suppressed by the weights, $w_{i}$, while the SN is inside the
overdensity. Removing SN~1993O from the dataset also removes the
$\chi^{2}$ step as well as the corresponding step in the attractor
mass. For the SSC location ($z=0.045$) variation of the SSC size or
the exclusion of SN~1993O has little effect on the fit quality as the
residual for SN~1993O is almost zero.

In order to better understand the limitations of our simplified model
we calculated the impact of inserting well-known major underdensities.
We first mimic the Local Void, then the Bo\"otes Void, finding only
5\% variations in the best fitting attractor mass and changes in
$\Delta\chi^2 \ll 1$. Likewise, older work based on the Abell Cluster
Catalog \citep{scaramella95}, or more modern works using very
carefully constructed X-ray selected cluster catalogs
\citep{kocevski07} reinforce the view that the SSC is the major
attractor within the redshift range of our supernova sample. Thus,
while simple, the model we employ is sufficient for the purposes of
exploring attractors in the direction of the SSC.

\begin{figure}[t]
   \includegraphics[width=3.6in]{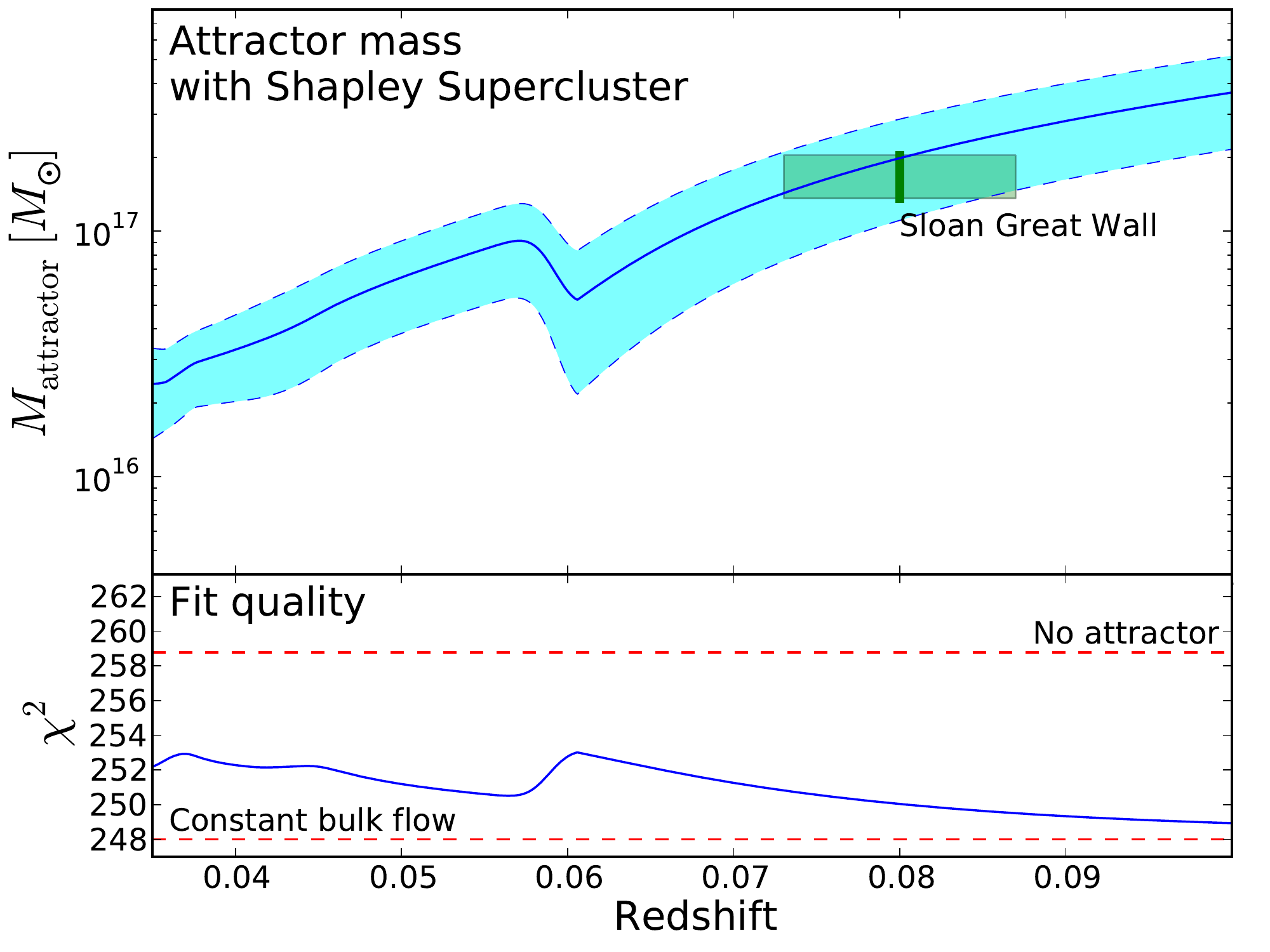}
   \caption{Attractor mass $M_{\textrm{attractor}}$ as a function of
     redshift $z$ as in Fig. \ref{ssc} where an attractor of SSC mass
     and size ($4.4\cdot 10^{16}M_{\sun},\ 50$~Mpc) was inserted at
     $z=0.045$. The green box shows the mass range as well as the
     approximate size of the Sloan Great Wall from
     \cite{2011MNRAS.417.2938S}. The red dotted line shows the
     $\chi^{2}$ value for a constant bulk flow.}
         \label{ssc-w-ssc}
   \end{figure}

To test for an attractor in addition to the SSC, we have repeated the
fit with the SSC inserted as a fixed overdensity of mass
$4.4\cdot10^{16}~M_{\sun}$ and radius 50~Mpc at $z=0.045$ (see Fig.
\ref{ssc-w-ssc}). For this mass fit the local minimum of the
$\chi^{2}$ value near the SSC location becomes shallower, while the
preference for a distant attractor remains. It was shown in
\cite{2011MNRAS.417.2938S} that a single supercluster of the mass of
the SSC can still be expected for a sphere of $200~h^{-1}$~Mpc radius.
An additional more distant supercluster seems an unlikely explanation,
since it would need to be rather massive and hence would be rare:
Extrapolating the mass function for superclusters from
\citet{2012arXiv1201.1382L}, we find that e.g. a factor of 3 more
massive cluster would already be $\sim50$ times rarer. A better
explanation might be non-collapsed structure. The Sloan Great Wall --
the largest known structure in the universe -- is located near the
direction of the SSC, at a redshift z=0.07-0.08. Its mass is estimated
to be $1.2 \cdot 10^{17} h^{-1} M_{\sun}$ by
\citet{2011MNRAS.417.2938S} and hence, together with the SSC, could
explain the size of the bulk flow. However, since it is an extended
source and not perfectly aligned with the SSC, our analysis cannot be
directly applied. A full analysis will be performed
elsewhere \citep{feindt13}.\\

\section{Conclusion \label{conc}}
We have investigated two independent samples of SNe~Ia --- the Union2
data as well as 117 new SNe from the \snf --- for anisotropies in the
expansion rate. The data were divided into redshift shells and for
each shell we obtain the degree of anisotropy by fitting for a bulk
velocity of the shell, on top of the expansion expected within a
baseline homogeneous isotropic universe with a flat $\Lambda$CDM
metric. Our key findings are:

\begin{itemize}
\item for the lowest redshift bin ($0.015<z<0.035$), bulk motion was
  detected at 2.3~$\sigma$ in the direction of the CMB dipole,
  consistent with previous findings. These constraints are dominated by
  the Union2 data.
\item for the intermediate redshift bin ($0.035<z<0.045$), bulk motion
  was observed with 1.1~$\sigma$ significance, consistent with the
  directions of the CMB dipole and the SSC. Only \snf and Union2 data
  combined show the observed weak evidence; by themselves the datasets
  show no significant trends.
\item for the next higher redshift bin ($0.045<z<0.06$), bulk motion
  was detected at 2.6~$\sigma$ significance in the direction of the
  CMB dipole analyzing \snf SNe. This reduces to 1.4 ~$\sigma$ if one
  combines them with the Union2 SNe. The constraints in this bin are
  driven by the \snf data.
\item we have shown that an attractor in the proximity of the SSC
    is only marginally consistent with our data and that the required
    attractor mass is larger than previous measurements of the SSC
    mass, suggesting the need for another gravitational source.
\item for the highest redshift bin ($0.06<z<0.1$) for which we can
  obtain meaningful constraints we find a bulk velocity that is
  consistent with zero, with a statistical uncertainty of around 400
  km~s$^{-1}$. Limiting the direction to that of the CMB dipole, the
  uncertainties are reduced to around 240 km~s$^{-1}$. Therefore we rule out
  (at $\sim 4~\sigma$) a Dark Flow as reported by \cite{kashlinsky10}
  (see below).
\item for all redshift shells analyzed, statistical errors dominate
  over systematic errors. The significance of the identified
  anisotropies was reproduced with an alternative analyses based on
  the smoothed residual method \cite{colin}.
\end{itemize}

\noindent The tentative observation, that the dipole motion does not
appear to converge at a distances scale of the SSC, i.e.\
150~$h^{-1}$Mpc, begs for an explanation. Within $\Lambda$CDM
structure formation one can expect bulk flows reaching to distances
beyond the ones observed here (see e.g. \citealt{nusser}), a
possibility
that will be explored in more detail elsewhere~\citep{feindt13}.\\
Other possible explanations for the observed anisotropies are related
to the CMB rest frame. \cite{wiltshire} argue in favor of differences
in the distances between the observer and the surface of last
scattering due to nearby foreground structures mimicking an additional
boost. More perplexing would be the presence of a Dark Flow, i.e. a
coherent flow extending throughout the visible Universe, possibly
originating in pre-inflation space-time \citep{kashlinsky08}. The case
for a Dark Flow is motivated by recent observation of coherent motion
of galaxy clusters, deduced from the kinematic Sunyaev-Zeldovich
effect \citep{kashlinsky10}. Bulk flow velocities of
$\sim1000$~km~s$^{-1}$ aligned with the CMB dipole were obtained for
shells with radius ranging from 270 Mpc to 800 Mpc. While it would be
another remarkable property of our Universe, their results rely on a
new method, and hence require confirmation. \cite{turnbull} already
derived an upper limit for a Dark Flow from modeling of the local
residual flow of $150 \pm 43~\textrm{km~s$^{-1}$}$ using SN and galaxy
data with a characteristic depth of $z=0.02$. As our highest redshift
bin overlaps with the lowest distance bin of the galaxy cluster
results of \citet{kashlinsky10}, we can now perform a
model-independent comparison. Our data show no evidence of large bulk
flows and are in a $\sim$4~$\sigma$ conflict with the claim that the
visible Universe is moving as a whole with a velocity of
$1000$~km~s$^{-1}$ relative to the CMB \citep{kashlinsky10}. While a
recent analyses of the \cite{planck_xiii} also rejects a Dark Flow,
that study relies on the kSZ effect like \cite{kashlinsky10}. Our
data, on the other hand, allow us to reject a Dark Flow based on an
independent method for the first time.

The redshift range from 0.03 to 0.08, for which the \snf has more than
doubled the number of available SNe, bridges the bulk flow
measurements obtained from individual galaxy distances to those
obtained from galaxy clusters. We have shown in this paper that SNe
are useful to address the outstanding questions about the motion of
the Local Group relative to the CMB. In this respect our conclusions
concerning the role of the SSC
should be considered tentative. However, with the ever growing number
of observed SNe~Ia from active and future surveys, combined with the
small systematic uncertainties associated with the measurements, it is
only a matter of time before these questions can be settled with
certainty.

\begin{acknowledgements}
  We are grateful to the technical and scientific staff of the
  University of Hawaii 2.2-meter telescope for their assistance in
  obtaining these data. D.~Birchall assisted with acquisition of the
  data presented here. We also thank the people of Hawaii for access
  to Mauna Kea. This work was supported in France by CNRS/IN2P3,
  CNRS/INSU, CNRS/PNC, and used the resources of the IN2P3 computer
  center; This work was supported by the DFG through TRR33 ``The Dark
  Universe'', and by National Natural Science Foundation of China
  (grant 10903010). U.\ Feindt acknowledges support by the
  Bonn-Cologne Graduate School of Physics and Astronomy. C.\ Wu
  acknowledges support from the National Natural Science Foundation of
  China grant 10903010. The IPNL collaborators acknowledge
  support from the Lyon Institute of Origins under grant
  ANR-10-LABX-66. This work was also supported by the Director, Office
  of Science, Office of High Energy and Nuclear Physics and the Office
  of Advanced Scientific Computing Research, of the U.S.\ Department
  of Energy (DOE) under Contract Nos. DE-FG02-92ER40704,
  DE-AC02-05CH11231, DE-FG02-06ER06-04, and DE-AC02-05CH11231; by a
  grant from the Gordon \& Betty Moore Foundation; by National Science
  Foundation Grant Nos. AST-0407297 (QUEST), and 0087344 \& 0426879
  (HPWREN); the France-Berkeley Fund; by an Explora'Doc Grant by the
  R\'egion Rh\^one-Alpes.
\end{acknowledgements}
\bibliographystyle{aa}
\bibliography{aniso}

\includepdf[pages=1-2]{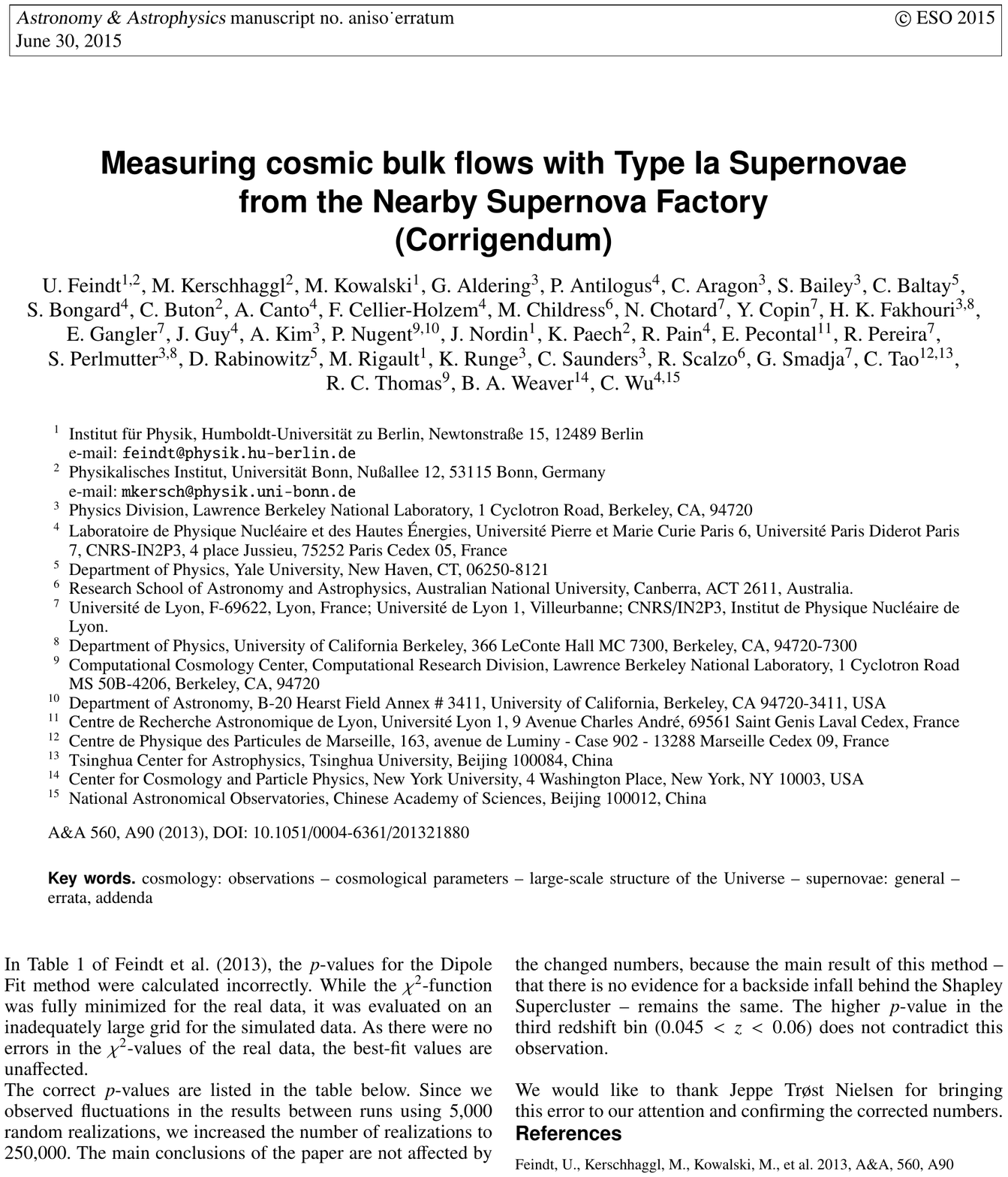}

\end{document}